\documentstyle[prl,aps,graphics,epsf]{revtex}
\begin{document}
\draft
\wideabs{
\title{Method of quantum computation with ``hot" trapped ions}
\author{Sara Schneider$^{1,2}$, Daniel F. V. James$^1$, 
and Gerard J. Milburn$^2$\\
\small{$^1$University of California, Los Alamos National Laboratory, Los 
Alamos, NM 87545, USA}\\
\small{$^2$Centre for Laser Science, Department of Physics,
The University of Queensland, St. Lucia, QLD 4072, Australia} }
\date{\today}
\maketitle
\begin{abstract}
We present a novel method of performing quantum logic gates in
trapped ion quantum computers
which does not require the ions to be cooled down to their vibrational 
center of mass (CM) mode ground state.
Our scheme employs adiabatic passages
and the conditional phase shift first investigated by D'Helon and
Milburn (C.~D'Helon and G.J.~Milburn, Phys. Rev. A {\bf 54}, 5141 (1996)).
\end{abstract}
\pacs{PACS numbers: 03.67.Lx, 42.50.Vk, 32.80.Pj}
}
\narrowtext
Not long ago it was recognized by a number of
workers that computations exploiting the 
quantum mechanical features of 
nature can perform efficiently certain tasks which 
are intractable with a classical computer 
\cite{Deutsch92,Shore94,Grover96}.
This discovery has motivated intensive 
research into apparatus which could be
used to perform quantum logic operations on single or multiple
quantum two-level systems (``qubits'').
Several possible physical implementations have been suggested. 
They include  bulk nuclear magnetic resonance (NMR) \cite{Gershenfeld97},
which was recently used to execute for the first time
a quantum algorithm \cite{Chuang98}.  However, there will
be very serious problems in implementing large scale
computations using high temperature bulk NMR \cite{Warren97}.  
A scalable low-temperature NMR device has been proposed
\cite{Kane98}, but there exist many formidable technological problems
to be overcome before even simple quantum logic operations
can be performed using it. Thus at the moment, ion trap 
quantum computation, first proposed by Cirac and Zoller \cite{Cirac95},
and demonstrated experimentally shortly afterwards \cite{NIST-gate},
is, arguably, the most promising quantum computation technology
for realizing systems of dozens of qubits in the foreseeable future.
An ion trap quantum computer consists of a string 
of ions in a linear radio-frequency trap. Two internal states of the ion 
compose each qubit and the center of mass (CM) 
vibrational mode of the ions' collective oscillations
acts as a quantum information bus, by means of which
quantum logic gate operations can be performed between pairs of ions. 

The most daunting technological problem to be overcome
in the realization of ion trap quantum computation is
the very fragile nature of the quantum mechanical ground-state
of the CM mode.
Cooling and maintaining the ions in this
ground state is required for performing logic gates 
in the manner proposed by Cirac and Zoller \cite{Cirac95}.
Any ``heating'' (i.e.~excitation by external fields) will diminish the
accuracy of a logic gate, thus leading to unreliable performance of 
the quantum computer as a whole, and maybe making the implementation
of even simple algorithms impossible. 
Eliminating all of the possible
causes of heating is a very demanding task.  Thus
it is desirable to investigate methods for performing quantum logic 
gates without the necessity of being in the ground state of 
the CM mode. 

Recently, three schemes to avoid the heating problem have been 
proposed: The first approach \cite{Poyatos97} very elegantly adapts 
ideas from atom interferometry. Consider two ions confined in a trap: 
Using a laser beam,
one of the ions can be given a state-dependent momentum kick, so
that it evolves in time into two spatially resolvable wavepackets,
corresponding to the two internal qubit states.
Because the second ion is strongly coupled to the first ion by the
Coulomb repulsion, it too will evolve into two wavepackets.
Since these wavepackets can be resolved using a laser, one of them
can be given a $\pi$-pulse dependent on the state of the first ion,
i.e.~a quantum gate can be realized.
One major problem with that scheme is that it is not 
scalable up to more than two ions, thus making it not very useful 
for quantum computation. 
Another potential difficulty is that it is advisable to use 
peculiar trapping potentials (e.g.~an axial confining potential 
$\sim |x|^{5/3}$ instead of the more usual $x^2$ harmonic confining 
potential).  A second scheme for quantum computation
has recently been proposed \cite{King98} (see also \cite{James98}). 
The ions oscillations have many modes other than the
center of mass mode  \cite{James97}.
As has been shown both theoretically and
experimentally \cite{King98,James98}, 
the heating rate of these ``higher'' modes 
is much smaller than that of the CM mode.
Thus these relatively more stable modes can be used as
a quantum bus instead of the unstable CM mode.
Drawbacks of this scheme are: the strength of the coupling of each ion 
to those higher order 
modes varies from ion to ion, making it harder to adjust the pulse 
durations; the higher modes are closer together in 
frequency space, thus making them harder to resolve; and also there is 
a `Debye-Waller' factor due to heating of the CM mode 
which alters the strength of the laser interaction by an 
undetermined factor. The third scheme \cite{Sorensen98}
which has been proposed very recently involves the use
of two laser beams driving transitions between virtual 
levels detuned from phonon resonances.

In this Letter we propose an entirely different approach to
the problem of performing quantum logic with `hot' trapped ions. 
We make use of the fact that although the ions are not 
necessarily in the CM mode ground state, they all share the 
same CM mode, thus enabling them to interact with each other. Our scheme
performs a controlled-rotation (CROT) gate between a pair of ions
designated control ($c$) and target ($t$). 
This gate operation consists of a condtional sign change which
takes place only if both ions are in the excited state. 
It can be realized by a sequence of four laser pulses, 
illustrated symbolically in figure 1.

We first do a conditional 
phase shift ${\cal{S}}_{t}$ between the target qubit and the 
CM phonon mode which changes the sign of the wavefunction
only if the
CM mode has an odd excitation and the target ion is in its excited state. 
This operation can be performed by applying a detuned laser pulse of well
defined duration with
the ion at the node of a standing wave of the addressing laser
\cite{D'Helon96}. 
The next step is 
to put an additional phonon into the CM mode, conditional on the state
of the control ion. 
This is realized by 
an adiabatic passage between the excited state $|1\rangle_{c}$ and some 
auxiliary state $|2\rangle_{c}$ of the control ion, 
which at the same time
puts a single phonon into the CM mode (see figure 2),
thereby changing an even phonon state to an odd phonon state (and 
vice-versa). 
The advantage of using adiabatic passage for this step is that the 
operation can be carried out {\em independent of the number of phonons}.
The next step is to perform a second conditional sign change
${\cal{S}}_{t}$.
Finally we disentangle the ion states from the 
CM mode by performing the adiabatic passage backwards.
As will be shown in detail below, these four pulses
produce the desired quantum logic gate {\em regardless
of the initial state of the phonon mode}. 
We will now describe those steps in detail. First let us consider
the various laser-ion interactions we will need.

To simplify our analysis we will assume that the CM
phonon mode is in a pure state given by the following
formula:
\begin{equation}
|\phi\rangle_{\mbox{\scriptsize CM}}= \sum_n a_n |n\rangle \,\, ,
\end{equation}
where $a_n$ are a set of unknown complex coefficients and
$|n\rangle$ is the Fock state of occupation number $n$.
It will be convienent in what follows to introduce the
odd and even parts of this wavefunction, viz.:
\begin{eqnarray}
|\mbox{even}\rangle_{\mbox{\scriptsize CM}}&=& \sum_n a_{2n} |2n\rangle \,\, ,
\nonumber \\
|\mbox{odd}\rangle_{\mbox{\scriptsize CM}}&=& \sum_n a_{2n+1} |2n+1\rangle \,\, .
\end{eqnarray}
We will also use the following notation for phonon
states to which a single quantum has been added:
\begin{eqnarray}
|\mbox{odd}^\prime\rangle_{\mbox{\scriptsize CM}}&=& 
\sum_n a_{2n} |2n+1\rangle \,\, , \nonumber \\
|\mbox{even}^\prime\rangle_{\mbox{\scriptsize CM}}&=& \sum_n a_{2n+1} |2n+2\rangle 
\,\, .
\end{eqnarray}

The conditional phase change between odd phonon number states and the 
excited internal state of an ion can be carried out using an effect 
first considered by D'Helon and Milburn \cite{D'Helon96}.  
They introduced a Hamiltonian 
for a two-level ion at the node of a detuned classical 
standing wave. 
In the limit of large detuning and for interaction times much 
greater than the vibrational period of the trap, this Hamiltonian 
for the $j$th ion is
\begin{equation}
H^{(j)} = \hbar a^\dagger a \chi (\sigma^{(j)}_z + 1/2) \,\, ,
\end{equation}
where $\sigma^{(j)}_z$ is the population inversion operator for the 
$j$th ion, $a$ and $a^\dagger$ are the annihilation and creation 
operators of the CM mode, and $\chi = \eta^2 
\Omega^2/(N\delta)$. Here 
$\eta$ is the Lamb-Dicke parameter, $\Omega$ is the Rabi 
frequency for the transitions between the two internal states of the 
ions, $N$ is the total number of ions and $\delta$ 
the detuning between the 
laser and the electronic transition. 
If we choose the duration $\tau$ of this interaction to be 
$\tau = \pi/\chi $, the time evolution is represented by the 
operator  
\begin{equation}
{\cal{S}}_{j} = \exp[- i a^\dagger a (\sigma^{(j)}_z + 1/2)\pi] \,\, .
\end{equation} 
This time evolution flips the phase of the ion
when the CM mode is in an odd state and the ion is in its 
excited state, thus providing us with a conditional phase shift 
for an ion and the CM mode. 

The adiabatic passage \cite{Bergmann95} which 
we require for our gate operation can be 
realized as follows:
We use two lasers, traditionally called the pump and 
the Stokes field. The 
pump laser is polarized to couple the qubit state $|1\rangle_c$ to 
some second auxiliary state
$|3\rangle_c$ and is detuned by an amount $\Delta$.
The Stokes laser couples to the red side band transition $|2\rangle_c
|n+1\rangle \leftrightarrow |3\rangle_c |n\rangle$, with the
same detuning $\Delta$. If the population we want to transfer 
adiabatically is 
initially in the state $|1\rangle_c |n\rangle$, 
we turn on the
Stokes field (i.e.~the sideband laser) and then slowly turn on 
the pump field (i.e.~the carrier laser) until both lasers are turned 
on fully. Then we slowly 
turn off the Stokes laser: this is the famous ``counter-intuitive'' 
pulse sequence used in adiabatic passage techniques \cite{Bergmann95}.
The adiabatic passage has to be performed very 
slowly. The condition in our scheme is that 
$T \gg 1/\Omega_{p,n}, 1/\Omega_{S,n}$, where $T$ is the duration of 
the adiabatic passage and $\Omega_{p,n}$ ($\Omega_{S,n}$) are the 
effective Rabi frequencies for the pump and the Stokes transition, 
respectively \cite{QM}.  
Using the adiabatic passage we can transfer the population from 
$|1\rangle_c |n\rangle$ to $|2\rangle_c |n+1\rangle$. To 
invert the adiabatic passage, we just have to interchange the roles 
of the pump and the Stokes field. We will denote the
adiabatic passage by operators ${\cal{A}}^{+}_{1}$ and ${\cal{A}}^{-}_{1}$
defined as follows:
\begin{eqnarray}
{\cal{A}}_j^+&:&|1\rangle_j |n\rangle \rightarrow  |2\rangle_j 
|n+1\rangle\nonumber \\
{\cal{A}}_j^-&:&|2\rangle_j |n+1\rangle \rightarrow  
|1\rangle_j|n\rangle\,\, .
\end{eqnarray}

Putting all those operations together in detail we can write down 
the step-by step states for our gate. 
We first perform the controlled phase shift between the target 
ion and the CM mode. Since this involves distinguishing 
even and odd CM mode states, we split them up in our representation,
as described above. 

\begin{eqnarray}
|0\rangle_c |0\rangle_t  \left\{|\mbox{even}\rangle_{\mbox{\scriptsize CM}}
+|\mbox{odd}\rangle_{\mbox{\scriptsize CM}}\right\} &  
\stackrel{{\cal{S}}_{t}}{\longrightarrow} & \nonumber \\
&  & \hspace*{-2.0cm}
|0\rangle_c |0\rangle_t \left\{|\mbox{even}\rangle_{\mbox{\scriptsize CM}}
+|\mbox{odd}\rangle_{\mbox{\scriptsize CM}}\right\} \nonumber  \\
|0\rangle_c |1\rangle_t \left\{|\mbox{even}\rangle_{\mbox{\scriptsize CM}}
+|\mbox{odd}\rangle_{\mbox{\scriptsize CM}}\right\}
& \stackrel{{\cal{S}}_{t}}{\longrightarrow} &  \nonumber  \\
&  & \hspace*{-2.0cm}
|0\rangle_c |1\rangle_t \left\{|\mbox{even}\rangle_{\mbox{\scriptsize CM}}
-|\mbox{odd}\rangle_{\mbox{\scriptsize CM}}\right\} \nonumber  \\
|1\rangle_c |0\rangle_t \left\{|\mbox{even}\rangle_{\mbox{\scriptsize CM}}
+|\mbox{odd}\rangle_{\mbox{\scriptsize CM}}\right\}
& \stackrel{{\cal{S}}_{t}}{\longrightarrow} & \nonumber  \\
&  & \hspace*{-2.0cm}
|1\rangle_c |0\rangle_t \left\{|\mbox{even}\rangle_{\mbox{\scriptsize CM}}
+|\mbox{odd}\rangle_{\mbox{\scriptsize CM}}\right\} \nonumber  \\
|1\rangle_c |1\rangle_t \left\{|\mbox{even}\rangle_{\mbox{\scriptsize CM}}
+|\mbox{odd}\rangle_{\mbox{\scriptsize CM}}\right\}
& \stackrel{{\cal{S}}_{t}}{\longrightarrow} & \nonumber  \\
&& \hspace*{-2.0cm}
|1\rangle_c |1\rangle_t \left\{|\mbox{even}\rangle_{\mbox{\scriptsize CM}}
-|\mbox{odd}\rangle_{\mbox{\scriptsize CM}}\right\}\,\,  .
\nonumber \\
\end{eqnarray}
The next step is the 
adiabatic passage as illustrated in fig. 2 and explained 
above. 
\begin{eqnarray}
|0\rangle_c |0\rangle_t \left\{|\mbox{even}\rangle_{\mbox{\scriptsize CM}}
+|\mbox{odd}\rangle_{\mbox{\scriptsize CM}}\right\}
& \stackrel{{\cal{A}}^{+}_{c}}{\longrightarrow}& \nonumber  \\
&  & \hspace*{-2.2cm}
|0\rangle_c |0\rangle_t \left\{|\mbox{even}\rangle_{\mbox{\scriptsize CM}}
+|\mbox{odd}\rangle_{\mbox{\scriptsize CM}}\right\} \nonumber  \\
|0\rangle_c |1\rangle_t \left\{|\mbox{even}\rangle_{\mbox{\scriptsize CM}}
-|\mbox{odd}\rangle_{\mbox{\scriptsize CM}}\right\}
&\stackrel{{\cal{A}}^{+}_{c}}{\longrightarrow}& \nonumber  \\ 
&  & \hspace*{-2.2cm}
|0\rangle_c |1\rangle_t \left\{|\mbox{even}\rangle_{\mbox{\scriptsize CM}}
-|\mbox{odd}\rangle_{\mbox{\scriptsize CM}}\right\} \nonumber  \\
|1\rangle_c |0\rangle_t \left\{|\mbox{even}\rangle_{\mbox{\scriptsize CM}}
+|\mbox{odd}\rangle_{\mbox{\scriptsize CM}}\right\}
&\stackrel{{\cal{A}}^{+}_{c}}{\longrightarrow} & \nonumber  \\ 
&  & \hspace*{-2.2cm}
|2\rangle_c |0\rangle_t \left\{|\mbox{odd}'\rangle_{\mbox{\scriptsize CM}}
+|\mbox{even}'\rangle_{\mbox{\scriptsize CM}}\right\} \nonumber  \\
|1\rangle_c |1\rangle_t \left\{|\mbox{even}\rangle_{\mbox{\scriptsize CM}}
-|\mbox{odd}\rangle_{\mbox{\scriptsize CM}}\right\}
& \stackrel{{\cal{A}}^{+}_{c}}{\longrightarrow} & \nonumber  \\
&  & \hspace*{-2.2cm}
|2\rangle_c |1\rangle_t \left\{|\mbox{odd}'\rangle_{\mbox{\scriptsize CM}}
-|\mbox{even}'\rangle_{\mbox{\scriptsize CM}}\right\} \, .
\nonumber \\
\end{eqnarray}
The next step is the conditional phase flip on  the target ion and the 
CM mode: 
\begin{eqnarray}
|0\rangle_c |0\rangle_t \left\{|\mbox{even}\rangle_{\mbox{\scriptsize CM}}
+|\mbox{odd}\rangle_{\mbox{\scriptsize CM}}\right\}
& \stackrel{{\cal{S}}_{t}}{\longrightarrow} & \nonumber  \\ 
&  & \hspace*{-2.7cm}
|0\rangle_c |0\rangle_t \left\{|\mbox{even}\rangle_{\mbox{\scriptsize CM}}
+|\mbox{odd}\rangle_{\mbox{\scriptsize CM}}\right\} \nonumber  \\
|0\rangle_c |1\rangle_t \left\{|\mbox{even}\rangle_{\mbox{\scriptsize CM}}
-|\mbox{odd}\rangle_{\mbox{\scriptsize CM}}\right\}
& \stackrel{{\cal{S}}_{t}}{\longrightarrow} & \nonumber  \\ 
&  & \hspace*{-2.7cm}
|0\rangle_c |1\rangle_t \left\{|\mbox{even}\rangle_{\mbox{\scriptsize CM}}
+|\mbox{odd}\rangle_{\mbox{\scriptsize CM}}\right\} \nonumber  \\
|2\rangle_c |0\rangle_t \left\{|\mbox{odd}'\rangle_{\mbox{\scriptsize CM}}
+|\mbox{even}'\rangle_{\mbox{\scriptsize CM}}\right\}
& \stackrel{{\cal{S}}_{t}}{\longrightarrow} & \nonumber  \\
&  & \hspace*{-2.7cm}
|2\rangle_c |0\rangle_t \left\{|\mbox{odd}'\rangle_{\mbox{\scriptsize CM}}
+|\mbox{even}'\rangle_{\mbox{\scriptsize CM}}\right\} \nonumber  \\
|2\rangle_c |1\rangle_t \left\{|\mbox{odd}'\rangle_{\mbox{\scriptsize CM}}
-|\mbox{even}'\rangle_{\mbox{\scriptsize CM}}\right\}
& \stackrel{{\cal{S}}_{t}}{\longrightarrow} & \nonumber  \\
&  & \hspace*{-2.7cm}
|2\rangle_c |1\rangle_t 
\left\{-|\mbox{odd}'\rangle_{\mbox{\scriptsize CM}}
-|\mbox{even}'\rangle_{\mbox{\scriptsize CM}}\right\}   .
\nonumber \\
\end{eqnarray}
The last step is the adiabatic passage backwards and the inversion of 
the rotation on the target ion: 
\begin{eqnarray}
|0\rangle_c |0\rangle_t \left\{|\mbox{even}\rangle_{\mbox{\scriptsize CM}}
+|\mbox{odd}\rangle_{\mbox{\scriptsize CM}}\right\}
& \stackrel{{\cal{A}}^{-}_{c}}{\longrightarrow} & \nonumber  \\ 
&  & \hspace*{-2.7cm}
|0\rangle_c |0\rangle_t \left\{|\mbox{even}\rangle_{\mbox{\scriptsize CM}}
+|\mbox{odd}\rangle_{\mbox{\scriptsize CM}}\right\} \nonumber  \\
|0\rangle_c |1\rangle_t \left\{|\mbox{even}\rangle_{\mbox{\scriptsize CM}}
+|\mbox{odd}\rangle_{\mbox{\scriptsize CM}}\right\}
& \stackrel{{\cal{A}}^{-}_{c}}{\longrightarrow} & \nonumber  \\
&  & \hspace*{-2.7cm}
|0\rangle_c |1\rangle_t \left\{|\mbox{even}\rangle_{\mbox{\scriptsize CM}}
+|\mbox{odd}\rangle_{\mbox{\scriptsize CM}}\right\} \nonumber  \\
|2\rangle_c |0\rangle_t \left\{|\mbox{odd}'\rangle_{\mbox{\scriptsize CM}}
+|\mbox{even}'\rangle_{\mbox{\scriptsize CM}}\right\}
& \stackrel{{\cal{A}}^{-}_{c}}{\longrightarrow} & \nonumber  \\ 
&  & \hspace*{-2.7cm}
|1\rangle_c |0\rangle_t \left\{|\mbox{even}\rangle_{\mbox{\scriptsize CM}}
+|\mbox{odd}\rangle_{\mbox{\scriptsize CM}}\right\} \nonumber  \\
|2\rangle_c |1\rangle_t \left\{-|\mbox{odd}'\rangle_{\mbox{\scriptsize CM}}
-|\mbox{even}'\rangle_{\mbox{\scriptsize CM}}\right\}
& \stackrel{{\cal{A}}^{-}_{c}}{\longrightarrow} & \nonumber  \\
&  & \hspace*{-2.7cm}
-|1\rangle_c |1\rangle_t \left\{|\mbox{even}\rangle_{\mbox{\scriptsize CM}}
+|\mbox{odd}\rangle_{\mbox{\scriptsize CM}}\right\}  .
\nonumber \\
\end{eqnarray}
Thus we end up with a controlled rotation gate between the ions
$c$ and $t$. A controlled-NOT (CNOT) gate can be realized by
performing $\pi/2$ rotation pulses on the target qubit
both before and after this series of operations.

For simplicity, we have analyzed these operations
under the assumption that the
state of the phonon CM mode can be described by an arbitrary
pure state.  More generally, one must assume that the CM mode is in
a mixed state, because it can be entangled with some unknown external
quantum system, for example the electromagnetic field causing the
heating. Provided we assume that this external system does
not become entangled with internal degrees of freedom of the qubits,
one can quite easily analyze the gate using a density matrix
formalism appropriate for mixed states. Since the 
adiabatic passage and the conditional phase shift all work for 
arbitrary CM mode phonon states, our principal result, that
gate operations can be performed between arbitrary pairs of qubits,
can be shown to be true under these circumstances.

A possible source of error in performing gate operations
using this scheme is the heating during gate operations.  
To perform logic operations, effectively
the quantum information stored in the two levels of the 
control qubit is transferred to the even and odd states of the CM mode. 
Heating mixes these two states, thereby degrading the information stored.
Since heating in ion traps is due to variuos sources of noise 
which have to be treated and modeled differently, this is a 
very involved problem which will be addressed in future work. 

This work was performed while one of us (S.S.) was
visiting Los Alamos National Laboratory;
she would like to thank Richard Hughes and the other members
of the quantum information team there for their hospitality.
She also acknowledges financial support from an University of Queensland 
Postgraduate Research Scholarship, from the Centre for Laser Science 
and from the Fellowship Fund - Branch of AFUW Qld.~Inc.
The authors would like to thank Ignacio Cirac, Richard Hughes, Brian King,
Paul Kwiat, Hideo Mabuchi, J{\"o}rg Steinbach, Gil Toombes 
and Dave Wineland for 
useful discussions and comments.
This work was supported by the U.S. National Security Agency
and the Australian Research Council International Program.

\newpage

\begin{figure}[!ht]
\vskip 1cm
\epsfxsize=8.5cm
\epsffile{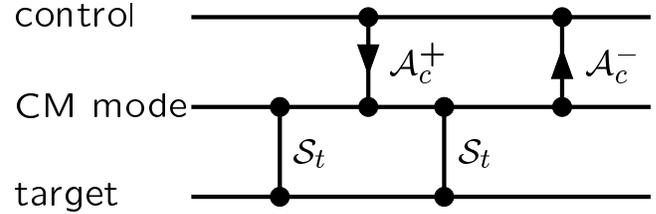}
\vskip 1cm
\caption{Schematic illustration of the steps
involved in the CROT gate with hot ions. The individual steps are
discussed in detail in the text.}
\label{figone}
\end{figure}

\begin{figure}[!ht]
\vskip 2cm
\epsfxsize=8.5cm
\epsffile{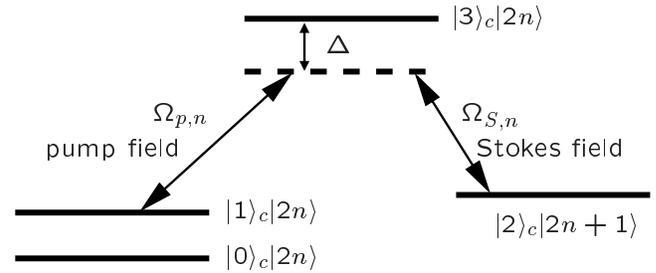}
\vskip 1cm
\caption{Schematic illustration of the level scheme of the
control ion used to realize the adiabatic passage operations
${\cal{A}}^{+}_{c}$ and ${\cal{A}}^{-}_{c}$.}
\label{figtwo}
\end{figure}

\end{document}